\documentclass[%
reprint,
superscriptaddress,
amsmath,amssymb,
aps,
prx,
twocolumn,
]{revtex4-2}

\usepackage{color}
\usepackage{amssymb}
\usepackage{braket}
\usepackage{graphicx}
\usepackage{dcolumn}
\usepackage{bm}
\usepackage{hyperref}
\usepackage{simplewick}
\usepackage{enumitem}
\usepackage{listings}
\usepackage{amsmath, amsthm}


\newcommand{\normdiamond}[1]{\left\|#1\right\|_\diamond}

\newtheorem{theorem}{Theorem}
\newtheorem{corollary}{Corollary}[theorem]
\newtheorem{lemma}[theorem]{Lemma}
\newtheorem{proposition}[theorem]{Proposition}

\usepackage{tikz}
\usepackage{circuitikz}
\usepackage{quantikz}
\usepackage{pgfgantt}
\usepackage{siunitx}

\definecolor{myred}{HTML}{FF1F5B}
\definecolor{myblue}{HTML}{009ADE}
\definecolor{mygreen}{HTML}{00CD6C}
\definecolor{myorange}{HTML}{F27522}
\colorlet{mucol}{red!90!black}
\colorlet{agg_col}{blue!80!}

\newcommand{\mytitle}{
  Programmable Dissipation via Partial Quantum Error Correction
}
\begin{document}

\title{
    \mytitle
}

\author{Sameer Dambal}
\affiliation{
Theoretical Division, Los Alamos National Laboratory, Los Alamos, NM, 87545, USA
}
\affiliation{Department of Physics, University of Houston, Houston, TX 77204, USA}

\author{Michael AD Taylor}
\affiliation{
Theoretical Division, Los Alamos National Laboratory, Los Alamos, NM, 87545, USA
}

\author{Yu Zhang}
\email{zhy@lanl.gov}
\affiliation{
Theoretical Division, Los Alamos National Laboratory, Los Alamos, NM, 87545, USA
}

\date{\today}

\begin{abstract}
Noise is typically treated as the adversary of quantum information processing. For open quantum dynamics, however, dissipation is part of the target physics, creating a tension with fault-tolerant architectures designed to suppress decoherence. Here we show that logical noise can instead be turned into a calibrated resource. We treat the error-correction cycle as a programmable primitive: one fault-tolerant round induces a logical completely positive trace-preserving map, and decoder/recovery randomization generates a controllable family of logical channels whose convex mixtures realize Kraus-channel mixing. This enables direct compilation of target dissipators into effective logical dynamics without explicit ancilla qubits for encoding the bath degree of freedoms. We derive an accuracy criterion for multi-step simulation in which the code distance is chosen so that uncontrolled logical errors remain a small fraction of the intended dissipation per step, rather than being driven below an arbitrarily small closed-system tolerance. Partial quantum error correction thus repurposes fault-tolerant structure to sculpt dissipation, offering a resource-efficient route to quantum simulation of open quantum systems.
\end{abstract}

\maketitle

\clearpage

{\bf Introduction}.
Noise has traditionally been the adversary of quantum computation. In the noisy intermediate-scale quantum (NISQ) era, this view has motivated shallow, noise-resilient circuit reduction methods~\cite{Alexeev:2025aa, Zhang:2022ab} and a broad class of error-mitigation techniques~\cite{RevModPhys.95.045005, Berg:2023aa} designed to extract useful signals before decoherence overwhelms the computation. As the field moves toward fault-tolerant quantum computation, the emphasis shifts from tolerating noise to actively suppressing it: quantum error correction (QEC) encodes logical information into redundant physical degrees of freedom, repeatedly extracts syndromes, and uses recovery operations to make the logical channel as close as possible to the intended unitary evolution~\cite{Shor:1995aa, Steane1996,KnillLaflamme1997, CalderbankShor1996, Laflamme1996Perfect, Dennis:2002aa, Fowler:2012aa, Terhal2015Quantum, Wang:2026aa, Andersen:2020aa, Krinner:2022aa, Acharya:2023aa}. This architecture is essential for high-fidelity quantum algorithms, but it is not naturally matched to many problems in quantum science. Real quantum systems are rarely perfectly isolated: dissipation and dephasing are not merely imperfections but often constitute the dynamics to be predicted, as in quantum transport, spin relaxation, non-equilibrium materials, quantum optical systems, and beyond~\cite{breuer2002theory, gardiner2004quantum, lindblad1976generators, Gorini1976GKLS, manzano2020short, mostame2012quantum, Dmitriev:2025aa, Lunghi:aa, Zhang:2013ab, Chen:2018aa}.

This mismatch creates a twofold overhead for simulating open quantum systems on fault-tolerant devices. First, many physical qubits are consumed to construct logical qubits. Second, the desired non-unitary reduced dynamics must be embedded back into a larger unitary evolution, for example through a Stinespring dilation, stochastic unraveling, postselection, block encoding, or related ancilla-based constructions~\cite{Kraus1983, Stinespring1955, nielsen2010quantum, WatrousBook18,Bauer:2020aa, Ollitrault:2021aa, Ding:2024aa, Schlimgen:2021vs, Hu:2020vx, Wang:2023aa, Delgado-Granados:2025aa}. 

Here we take a different route: we propose partial QEC as a logical dissipative primitive. A syndrome-extraction cycle is a physical completely positive trace-preserving (CPTP) process once measurement outcomes are averaged over and the classical record is discarded. Encoding and decoding convert this physical process into an effective logical CPTP map. Moreover, if the recovery operation, decoder tie breaking, logical-frame update, or a small recovery gadget is sampled from a classical distribution, the resulting logical channel is a convex combination of implementable logical channels. The partial QEC proposed in this work uses this freedom not merely to preserve information, as in approximate or noise-adapted QEC~\cite{Leung1997ApproxQEC, BenyOreshkov2010, Kribs2005OQEC}, but to synthesize a target dissipative map directly at the logical level.

We propose two complementary strategies to achieve the goal. 
{\it Strategy A} is model-aware: the calibrated physical noise, syndrome circuit, decoder, and randomized recovery library are propagated to a family of logical maps~\cite{Berg:2023aa}. The target dissipator is then fitted within the convex hull of that family. 
Logical noise aligned with the target dissipator is no longer counted as failure, so the code distance is set by the residual wrong-channel errors, fitting error and model mismatch.
{\it Strategy B} is post-correction programming: the physical noise is first removed by conventional QEC, and a logical CPTP map is then sampled from an additional programming library, like ancilla-enabled corrected logical operations~\cite{Reinhold:2020aa}.  This route is model-agnostic with respect to the physical-to-logical mapping, and it can implement nonunitary logical dynamics without allocating explicit bath registers for a Stinespring dilation. 
Strategy B therefore does not reduce the code distance required for the correction stage, but it can reduce the channel-implementation overhead relative to dilation-based simulation.

{\bf Results and Discussion}.
For a system $\mathcal{S}$ interacting with an environment $\mathcal{E}$, the total Hamiltonian $H_T$ is given by
$H_T = H_\mathcal{S} + H_\mathcal{E} + H_\mathcal{SE}$. With $\hbar=1$, the dynamics of the reduced system can be obtained by tracing out the environmental DoF from the unitary dynamics of the entire system, i.e.,
\begin{equation}
 i \frac{\partial \rho_{\mathcal{S}}(t)}{\partial t} =\text{Tr}_{\mathcal{E}} [H_T, \rho_T].
\end{equation}
For an initially factorized preparation $\rho_T(0)=\rho_{\mathcal S}(0)\otimes\rho_{\mathcal E}$, the dynamics of the system is given by 
\begin{equation}
\rho_{\mathcal S}(t)=\operatorname{Tr}_{\mathcal E} \left[U(t)\bigl(\rho_{\mathcal S}(0)\otimes\rho_{\mathcal E}\bigr)U^\dagger(t)\right]
=\mathcal E_{\rm tar}^{(t)}(\rho_{\mathcal S}(0)),
\end{equation}
where $U = e^{-iH_T t}$. 
For a fixed preparation of the environment, this map is completely positive and trace-preserving (CPTP) and admits a Kraus representation~\cite{Kraus1983},
\begin{equation}
  \rho_{\mathcal{S}}(t) =
  \sum_{i} K_i \rho_{\mathcal{S}} K_i^\dagger,
  \quad \sum_i K_i^\dagger K_i=\mathbb I.
\end{equation}
For example, for the widely used Lindbladian dynamics, its corresponding Kraus formalism is $ K_0 = \mathbb{I} - (iH + \frac{1}{2}\sum_i L^\dag_i L_i)\Delta t, \quad K_i = \sqrt{\Delta t}L_i$.

In fault-tolerant architectures, however, the elementary cycle is not the target dissipative dynamics $\mathcal{E}_{\mathrm{tar}}^{(\tau)}$ but a quantum error correction (QEC)~\cite{Devitt2013} round: syndrome extraction ($\mathcal{S}$), decoding ($\mathcal{D}$), and conditional recovery $\mathcal{R}$. This QEC round naturally defines a CPTP map on the \emph{encoded} degrees of freedom (DoF) when one averages over all syndrome outcomes.
The central idea of partial QEC is to use this map not only to suppress errors, but also to shape the effective logical dynamics so that inevitable logical noise becomes the intended dissipation.

To formalize this perspective, let $\mathcal{C}_d$ denote an ideal encoding map into a distance-$d$ code space,
and let a single fault-tolerant round (including physical noise, syndrome extraction, and conditional recovery) define an
instrument $\{\mathcal{M}_s\}$ followed by recovery channels $\{\mathcal{A}_s\}$. The syndrome-averaged channel on the
encoded block is
\begin{equation}
\mathcal{Q}_d =\sum_{s} \mathcal{A}_s\circ \mathcal{M}_s,
\label{eq:Qd_def}
\end{equation}
where  $\circ$ denotes composition of maps (channels); for instance, $(\mathcal{D}\circ \mathcal{U})(\rho)=\mathcal{D}(U\rho U^\dagger)$. The induced logical channel (in an ideal decode picture) is
\begin{equation}
\mathcal{B}_d = \mathcal{D}_d\circ \mathcal{Q}_d\circ \mathcal{C}_d,
\label{eq:Bd_def}
\end{equation}
where $\mathcal{D}_d$ is an ideal decoding map back to the logical Hilbert space~\footnote{Equivalently one may stay entirely within the code space and use logical Pauli/frame tracking; the abstract channel $\mathcal{B}_d$ captures the logical input--output relation after discarding classical records.}. However, we will show (prove) that one QEC round can fundamentally generate a CPTP logical channel.

\begin{lemma}[One QEC round induces a CPTP logical channel]
\label{lem:QEC_round_CPTP}
Let $D$ denote the data block (the encoded physical qubits) and $A$ the syndrome ancillas. Assume that in each QEC round:
(i) ancillas are prepared in a fixed state $\omega_A$ (freshly prepared or reset),
(ii) a fixed noisy syndrome-extraction circuit is applied (modeled as a CPTP channel on $DA$),
(iii) ancillas are measured with outcomes $s$ and we do \emph{not} postselect (we sum over all $s$),
(iv) conditioned on $s$, a recovery channel $\mathcal{A}_s$ (TP) is applied to $D$,
(v) ancillas and classical records are discarded.
Then the syndrome-averaged map (Eq.~\eqref{eq:Qd_def}) on the data block, 
i.e., $\mathcal{Q}_d(\rho_D)=\sum_s \mathcal{A}_s \left(\mathcal{M}_s(\rho_D)\right)$, 
is CPTP. Moreover, if $\mathcal{C}_d$ is an encoding CPTP map and $\mathcal{D}_d$ is a decoding CPTP map (on the logical subsystem), then the induced logical map $\mathcal{B}_d$ in Eq.~\eqref{eq:Bd_def} is CPTP.
\end{lemma}

Lemma~\ref{lem:QEC_round_CPTP} (see the Supplementary Material (SM) for the proof) indicates the programmability of the dissipation channels of logical qubits via controlling the recovery/decoder randomization. 
The key control lever is that QEC does not specify a unique channel: by altering the decoding rule, adding logical frame updates, or inserting small logical gadgets (including ancilla-assisted steps with reset), we obtain a \emph{family} of implementable logical CPTP maps per round. 
Let $\{\mathcal{G}_k\}_{k=1}^M$ be such a family, each realizable with the same code distance $d$ and comparable physical depth. If we sample $k$ with probabilities $r_k$ and discard the classical key, the effective programmed channel is
\begin{equation}
\mathcal{G}_r
=\sum_{k=1}^{M} r_k\mathcal{G}_k,
\label{eq:Gr_def}
\end{equation}
where $r_k\ge 0, \sum_k r_k=1$. Eq.~\ref{eq:Gr_def} is precisely the structure of a Kraus map: convex mixing of CP evolutions is itself CPTP. This leads to the second elementary ingredient of this work as follows:
\begin{lemma}[Randomized recovery mixes Kraus channels]
\label{lem:randomization_kraus_mixing}
Let $\{\mathcal{G}_k\}_{k=1}^{M}$ be implementable logical CPTP maps with Kraus forms $\mathcal{G}_k(\rho)=\sum_\alpha G_{k,\alpha}\rho G_{k,\alpha}^{\dagger}$.  For any probability vector $r\in\Delta_M \equiv\{r\in\mathbb{R}^M: r_k\ge 0,\ \sum_k r_k=1\}.$, the sampled map
\begin{equation}
\mathcal{G}_r=\sum_{k=1}^{M}r_k\mathcal{G}_k
\label{eq:randomized_channel_main}
\end{equation}
is CPTP and admits the Kraus representation
\begin{equation}
\mathcal{G}_r(\rho)=\sum_{k=1}^M\sum_{\alpha=1}^{K_k} \widetilde{G}_{k,\alpha}\rho \widetilde{G}_{k,\alpha}^\dagger,
\quad \widetilde{G}_{k,\alpha}:=\sqrt{r_k} G_{k,\alpha}.
\label{eq:Gr_kraus}
\end{equation}
\end{lemma}

With the programmable logical CPTP channels in Lemma~\ref{lem:randomization_kraus_mixing} (see the SM for the proof), then it's possible to tune the logical channels to mimic arbitrary target open quantum system dissipation channels. This is equivalent to a \emph{logical open quantum system compilation problem.} We target a per-step logical evolution of the form
\begin{equation}
\mathcal{E}_{\mathrm{tar}}^{(\tau)}
\approx
\mathcal{D}_{\mathrm{tar}}^{(\tau)}\circ \mathcal{U}^{(\tau)},
\label{eq:target_map}
\end{equation}
where $\mathcal{U}^{(\tau)}(\rho)=U\rho U^\dagger$ implements coherent dynamics ($U=e^{-iH\tau}$) and $\mathcal{D}_{\mathrm{tar}}^{(\tau)}=e^{\mathcal{L}_D}$ captures dissipation.
On encoded hardware, we implement $\mathcal{U}^{(\tau)}$ fault-tolerantly at the logical level, while the dissipative part is realized by \emph{programming} the logical noise through partial QEC. A minimal architecture for one stroboscopic step is
\begin{equation}
\mathcal{E}_{\mathrm{sim}}^{(\tau)} = \mathcal{G}_r \circ \mathcal{B}_d \circ \mathcal{U}^{(\tau)}.
\label{eq:sim_architecture}
\end{equation}
For Strategy A, a physical noise model $\theta$ and a recovery choice $k$ define
\begin{equation}
\mathcal{B}_{d,\theta}^{(k)}=\mathcal{D}_d\circ\left(\sum_s\mathcal{A}_s^{(k)}\circ\mathcal{M}_{s,\theta}\right)\circ\mathcal{C}_d .
\label{eq:strategyA_library_main}
\end{equation}
Sampling $k$ with probabilities $r_k$ gives
\begin{equation}
\mathcal{E}_{\mathrm{sim}, A}^{(\tau)}=\sum_{k=1}^{M_A} r_k \mathcal{B}_{d,\theta}^{(k)}\circ \mathcal{U}^{(\tau)}.
\label{eq:main_A}
\end{equation}
For Strategy B, a corrected baseline logical channel $\mathcal{B}_d^{\rm FT}$ is followed by a sampled logical programming map $\mathcal{G}_s=\sum_{\ell=1}^{M_B}s_\ell\mathcal{G}_\ell$:
\begin{equation}
\mathcal{E}_{\mathrm{sim}, B}^{(\tau)}=\mathcal{G}_s\circ \mathcal{B}_d^{\rm FT}\circ \mathcal{U}^{(\tau)}.
\label{eq:main_B}
\end{equation}
Then the compilation task is to choose $r$ ($s$ for Strategy B) and $d$ so that $\mathcal{E}_{\mathrm{sim}}^{(\tau)}$ approximates $\mathcal{E}_{\mathrm{tar}}^{(\tau)}$ to a prescribed accuracy, while minimizing physical qubits.

To achieve this goal, here we turn to a more convenient representation, i.e., the Choi matrix~\cite{choi1975completely} $J(\mathcal{E})=(\mathcal{E}\otimes \mathcal{I})(\ket{\Phi} \bra{\Phi})$, for $\ket{\Phi}$ maximally entangled. Since $J$ is linear in $\mathcal{E}$, Eq.~\eqref{eq:Gr_def} implies that
\begin{align}
J(\mathcal{G}_r)= & \sum_k r_k J(\mathcal{G}_k), \\
J(\mathcal{E}_{\mathrm{sim}}^{(\tau)})=& \left(\sum_k r_k J(\mathcal{G}_k)\right)\circledast J(\mathcal{B}_d\circ\mathcal{U}^{(\tau)}),
\label{eq:choi_affine}
\end{align}
where $\circledast$ denotes the bilinear composition operation in the Choi picture. Thus, finding $r$ is a convex feasibility problem: $r$ lives on a simplex and the objective (e.g.\ diamond distance to target) is convex in $r$. Here, we then can prove that the target dissipative channel can be reached via partial QEC.

\begin{theorem}[Convex reachability by partial QEC and universal logical open quantum system compilation]
\label{thm:reachability_convex}
For a fixed code distance $d$ and an implementable family of logical channels $\{\mathcal{G}_k\}_{k=1}^M$,
the Strategy-A reachable set
\begin{equation}
\mathfrak{C}_{A,d}(\theta)=\left\{\sum_{k=1}^{M_A} r_k \mathcal{B}_{d,\theta}^{(k)}\circ\mathcal{U}^{(\tau)}:r\in\Delta_{M_A}\right\}
\label{eq:strategyA_reachable_set}
\end{equation}
and the Strategy-B reachable set
\begin{equation}
\mathfrak{C}_{B,d}=\left\{\sum_{\ell=1}^{M_B} s_\ell \mathcal{G}_\ell\circ\mathcal{B}_d^{\rm FT}\circ\mathcal{U}^{(\tau)}:s\in\Delta_{M_B}\right\}
\label{eq:strategyB_reachable_set}
\end{equation}
are convex subsets of logical channel space.
Their Choi matrices are affine in the probabilities $r$ and $s$.  Hence exact reachability of $\mathcal{E}_{\rm tar}^{(\tau)}$ is a linear feasibility problem, while best approximation, $\min_{r\in\Delta_M}\|\mathcal{E}_{\mathrm{sim}}^{(\tau)}-\mathcal{E}_{\mathrm{tar}}^{(\tau)}\|_\diamond$,  in any convex channel norm is a convex optimization over the corresponding simplex.

Moreover, any finite-dimensional Markovian or non-Markovian open quantum system time step is a CPTP map on the logical Hilbert space.  Therefore, if the Strategy-A physical-to-logical library or the Strategy-B logical programming library has a convex hull that contains $\mathcal{E}_{\rm tar}^{(\tau)}$ after the coherent part in Eq.~\eqref{eq:target_map}, partial QEC can reproduce that time step exactly.  
\end{theorem}

This theorem (see the SM for the proof) is the formal reason that partial QEC can mimic arbitrary open quantum system dynamics once a sufficiently expressive logical CPTP library is available.  A Pauli-frame library is universal for Pauli channels such as pure dephasing, but not for non-unital channels.  Amplitude damping requires calibrated physical relaxation in Strategy A or logical measurement/reset/feedback primitives in Strategy B.

The two strategies differ in the error budget. Let
\begin{equation}
\Delta_{\rm tar}(\tau)=\normdiamond{\mathcal{D}_{\rm tar}^{(\tau)}-\mathcal{I}},
\label{eq:Delta_tar_main}
\end{equation}
which is $O(\gamma\tau)$ for weak Markovian dissipation with a rate of $\gamma$.

\begin{theorem}[Distance selection for open quantum system simulation]
\label{thm:distance_selection}
Suppose a simulation uses $m=T/\tau$ steps, if the per-step diamond-norm error satisfies $\normdiamond{\mathcal{E}_{\rm sim}^{(\tau)}-\mathcal{E}_{\rm tar}^{(\tau)}}\le\epsilon_{\rm step}$, then
\begin{equation}
\normdiamond{(\mathcal{E}_{\rm sim}^{(\tau)})^m-(\mathcal{E}_{\rm tar}^{(\tau)})^m}
\le m\epsilon_{\rm step} .
\label{eq:multi_step_bound_main}
\end{equation}
Consequently, total error $\varepsilon$ is guaranteed by $\epsilon_{\rm step}\le\varepsilon/m$.  In Strategy A, let $\epsilon_{\rm un}(d)$ be the component of the modeled logical channel outside the target-compatible dissipative family, let $\epsilon_{\rm prog}^{A}(d,\theta)$ be the optimal channel-fitting error, and let $\nu_d$ bound physical-model mismatch.  A sufficient target-aware distance rule is
\begin{equation}
\epsilon_{\rm un}(d)+\epsilon_{\rm prog}^{A}(d,\theta)+\nu_d
\lesssim \min\left\{\frac{\varepsilon}{m}, \zeta\Delta_{\rm tar}(\tau)\right\},
\qquad \zeta\ll1.
\label{eq:strategyA_distance_rule_main}
\end{equation}
In Strategy B, the conventional correction stage must still approximate the identity before the desired logical CPTP map is added, so the corresponding sufficient rule is
\begin{equation}
\epsilon_{\rm FT}(d)+\epsilon_{\rm prog}^{B}(d)\lesssim \frac{\varepsilon}{m}.
\label{eq:strategyB_distance_rule_main}
\end{equation}
Thus Strategy A can reduce the code distance when useful physical noise is available, while Strategy B primarily removes the need for explicit environment registers or channel dilations.
\end{theorem}

To translate this condition into qubit overhead, use the standard below-threshold scaling ansatz $p_L(d)=A\left(\frac{p_{\rm phys}}{p_{\rm th}}\right)^{(d+1)/2}, \quad p_{\rm phys}<p_{\rm th}$
for the unwanted logical component.  If $Cp_L(d)\le x$ ($x\in\{\delta, \eta\epsilon\}$), then the odd code distance is approximated by $d(x)\simeq 2\frac{\log[x/(AC)]}{\log(p_{\rm phys}/p_{\rm th})}-1$, rounded up to the next permitted distance.

\begin{corollary}[Target-aware qubit savings are specific to Strategy A]
\label{cor:qubit_savings}
For $n_L$ logical qubits encoded in rotated surface-code patches with $N_{\rm patch}(d)\simeq2d^2-1$, the footprint is $N_{\rm tot}(d)\simeq n_L(2d^2-1)$ up to routing and lattice-surgery overheads.  If Strategy A can use the target-aware threshold $x_A=\zeta\Delta_{\rm tar}(\tau)$ while full QEC or Strategy B uses the closed-system threshold $x_B=\varepsilon/m$, then
\begin{equation}
\frac{N_{\rm tot}(d_B)}{N_{\rm tot}(d_A)}
\simeq
\frac{2d(x_B)^2-1}{2d(x_A)^2-1}
\simeq \left[\frac{d(x_B)}{d(x_A)}\right]^2 .
\label{eq:savings_factor}
\end{equation}
Substantial code-distance savings require $x_A\gg x_B$, i.e. the intended dissipative change per step is much larger than the closed-system logical-error tolerance.  Strategy B does not provide this code-distance saving, but it can still reduce the resources otherwise spent on environment registers, controlled dilations and postselection.
\end{corollary}

{\bf Scaling of searching for $r_k$ within partial QEC}.
Theorem~\ref{thm:reachability_convex} shows that the search over \(r\) remains a finite-dimensional convex problem: its cost is controlled not by the exponentially large physical Hilbert space, but by the size \(M(d)\) of the programmable library and by the best reachable error at distance \(d\) because we are searching in the logical space. Let \(D_L=2^{k_L}\) be the logical Hilbert-space dimension. Then, the Hermitian Choi matrices on the logical channel space have \(D_L^4\) real parameters, while the trace-preserving condition imposes \(D_L^2\) independent linear constraints. Consequently, the set of Hermitian TP Choi matrices has a real affine dimension of $n_{\mathrm{aff}}:=D^4_L - D^2_L$.  
Choose any Euclidean coordinates on this affine space, and let \(x_k^{(d)}\) and \(x_{\mathrm{tar}}\) denote the coordinate vectors of \(J(\mathcal{G}_k\circ \mathcal{B}_d\circ \mathcal{U}^{(\tau)})\) and \(J(\mathcal{E}_{\mathrm{tar}}^{(\tau)})\), respectively. 

Define $R_d:=\max_{1\le k\le M(d)}\|x_k^{(d)}\|_2$, which is the radius of the programmable space in its centered coordinate representation. For a fixed logical space, $R_d$ is $\mathcal{O}(1)$. For any probability vector $r\in\Delta_{M(d)}$, $S_r:=\{k: r_k>0\}$, i.e., $|S_r|$ defines the number of active channels in the mixture $\mathcal{G}_r=\sum_{k=1}^{M(d)} r_k \mathcal{G}_k$. 
Define the best reachable per-step deviation at distance \(d\) by
\begin{equation}
\delta_\star(d) := \min_{r\in\Delta_{M(d)}}
\left\| \sum_{k=1}^{M(d)} r_k x_k^{(d)} - x_{\mathrm{tar}} \right\|_2 \leq \eta.
\label{eq:delta_from_search}
\end{equation}

\begin{proposition}[Scaling of searching for mixing weights $\{r_k\}$]
\label{prop:scaling_rk_search}
Let $M(d)$ be the number of implementable logical channels in the programmable family at code distance $d$ and $\delta_\star(d)$ be defined by Eq.~\eqref{eq:delta_from_search}, then:

\begin{enumerate} 
    \item If $\eta=0$, the exact solution $r_k \in\Delta_{M(d)}$, i.e., $\sum_{k=1}^{M(d)} r_k x_k^{(d)} = x_{\mathrm{tar}}$, can be found with $S_r$ bounded by $|S_r| \leq n_{\mathrm{aff}}$.

    \item The projection-free Frank-Wolfe algorithm finds the optimal solution $r_k$ for the objective function over the simplex, with a convergence rate scaling as \cite{pokutta2024frank} (see the SM for the proof):
    \begin{equation}
        \mathcal{O}\left({\frac{LM^2(d)}{\eta}}\right).
    \end{equation}

    where $L$ is the Lipschitz coefficient of the objective function \eqref{eq:delta_from_search}, $M(d)$ is the dimension of the simplex as defined earlier, and $\eta$ is the accuracy.

\end{enumerate}

\end{proposition}

Theorems~\ref{thm:reachability_convex},~\ref{thm:distance_selection} and Proposition~\ref{prop:scaling_rk_search} constitute the main results of this work.
They show that partial QEC is not simply an imperfect approximation to full QEC, but a programmable mechanism for converting native logical noise into useful open quantum system dynamics.  Theorem~\ref{thm:reachability_convex} identifies the convex family of dissipative logical channels reachable by programmable recovery, while Theorem~\ref{thm:distance_selection} gives a target-aware distance rule: only the component of the logical noise outside the target-compatible dissipative family must be strongly suppressed.  Thus, in Strategy~A, noise aligned with the desired generator is retained as a resource, enabling code-distance and qubit-footprint reductions whenever $\Delta_{\rm tar}(\tau)$ exceeds the closed-system tolerance scale $\varepsilon/m$.  Proposition~\ref{prop:scaling_rk_search} further shows that the required mixing weights can be found by a finite-dimensional convex search in logical channel space, with complexity controlled by the programmable library size $M(d)$ rather than the physical Hilbert-space dimension.  Together, these results provide a quantitative framework for target-aware fault tolerance in open quantum system quantum simulation, where the code and recovery are designed to suppress the wrong noise while preserving and programming the useful part.

{\bf Numerical examples.}
A natural target application of the proposed method is spin (or excitonic) dynamics.  For two sites encoded as spin-$1/2$ degrees of freedom, we use
\begin{equation}
H=-\frac{\epsilon_1}{2}Z_1-\frac{\epsilon_2}{2}Z_2+\frac{J}{2}(X_1X_2+Y_1Y_2),
\end{equation}
which reduces to the standard hopping Hamiltonian in the single-excitation subspace.  A minimal dissipator contains local dephasing, correlated dephasing and optional loss,
\begin{align}
\dot\rho ={}& -i[H,\rho]
+\sum_{j=1}^2\frac{\gamma_j}{2}(Z_j\rho Z_j-\rho)
+\frac{\gamma_{12}}{2}(Z_1Z_2\rho Z_1Z_2-\rho)\nonumber\\
&+\sum_{j=1}^2\kappa_j\left(\sigma_j^-\rho\sigma_j^+-\frac{1}{2}\{\sigma^+_j\sigma^{-}_j,\rho\}\right).
\label{eq:exciton_main}
\end{align}
The dephasing part is exactly a Pauli channel.  For one qubit, $\mathcal{D}_Z^{(\tau)}(\rho)=\frac{1+e^{-\gamma\tau}}{2}\rho+\frac{1-e^{-\gamma\tau}}{2}Z\rho Z$.
So logical $Z$-frame sampling directly programs the dissipator.  Amplitude damping is different: it is non-unital, with Kraus operators $K_0=\ket{0} \bra{0}+\sqrt{1-\lambda}\ket{1} \bra{1}$ and $K_1=\sqrt{\lambda}\ket{0} \bra{1}$, and hence cannot be generated by Pauli-frame randomization alone.  It can enter Strategy A through calibrated physical relaxation, or Strategy B through logical reset/feedback primitives.

\begin{figure}[htb]
\centering
\includegraphics[width=\linewidth]{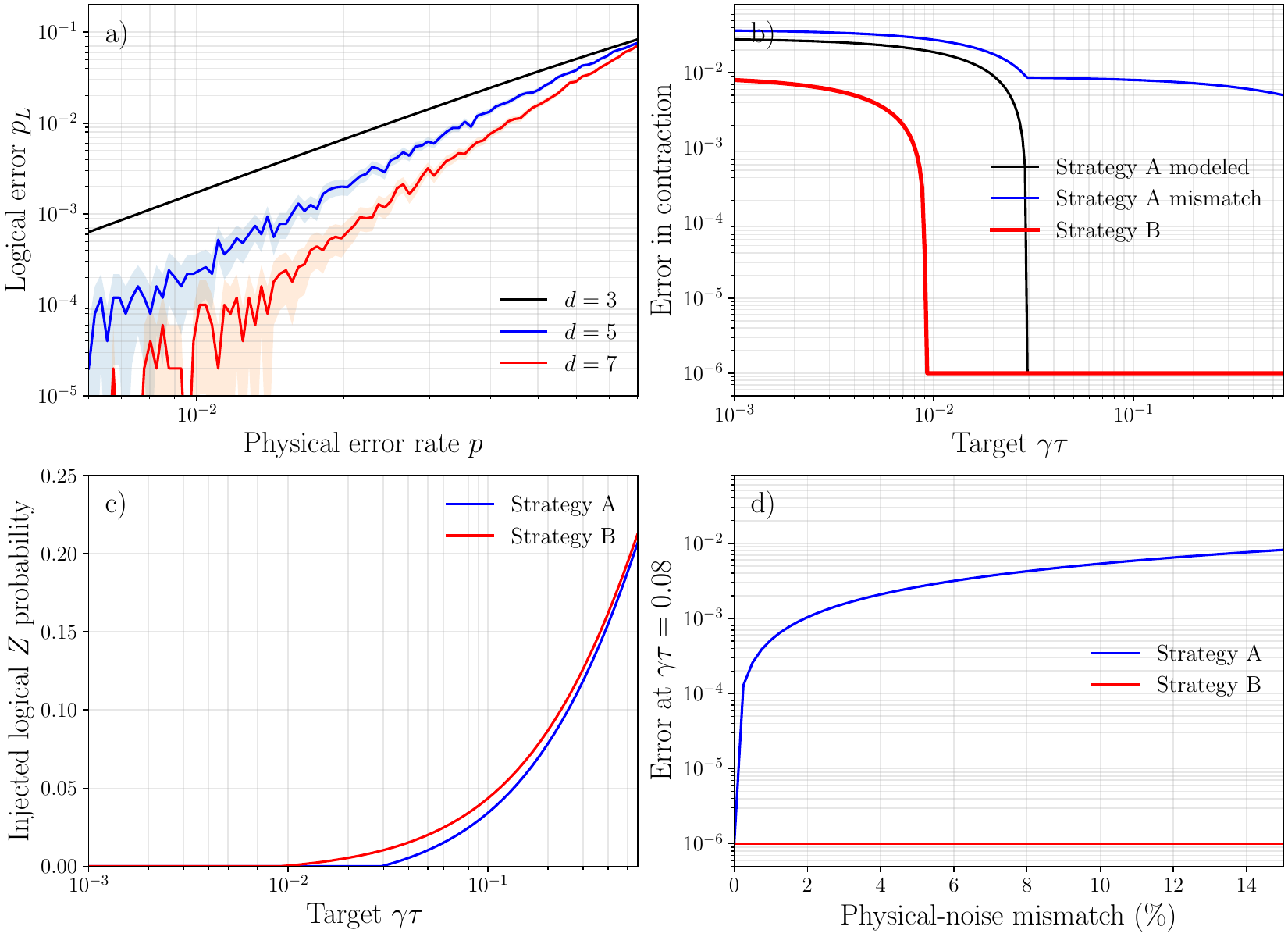}
\caption{
{\bf Logical channel compilation}: a) Logical error extracted from an explicit surface-code stabilizer workflow: physical error patterns are enumerated via a minimum weight perfect matching (MWPM) decoder~\cite{higgott2021:mwpm};
b)  Channel fitting error for a one-qubit dephasing target; To illustrate two different strategies, here Strategy A uses the modeled $d=3$ extracted logical channel as part of the target dissipator, so the fit is exact when the assumed noise model is feasible; a 15\% physical-noise mismatch leaves a calibrated residual. Strategy B corrects more strongly first ($d=7$ here) and then programs the dephasing. c) logical $Z$-frame probabilities; d)  Sensitivity at fixed target strength $\gamma\tau=0.08$. Strategy A is sensitive to noise-model mismatch because it uses physical noise as a resource, whereas Strategy B is insensitive to this particular mismatch because it re-fits the logical channel after correction.}
\label{fig:channel_compilation} 
\end{figure}

Figure~\ref{fig:channel_compilation} demonstrates the logical-channel compilation primitive underlying the partial QEC. Rather than assuming an logical noise rate, we extract the effective logical channel from an explicit stabilizer-syndrome and decoder calculation (Figure~\ref{fig:channel_compilation}a), express it in the Pauli-transfer matrix (PTM) representation, and then choose the randomized recovery weights $r_k$ to match a target dephasing channel. This directly illustrates the main theorem: after encoding, syndrome processing and recovery, a QEC round is itself a programmable logical CPTP map, and classical randomization turns the accessible logical maps into a convex set. The fitting problem is therefore not a search over the exponentially large physical Hilbert space, but a convex search in the logical Choi space.

The comparison between Strategies A and B highlights the central trade-off. Strategy A uses the calibrated native logical dissipation as part of the simulated channel, so it requires less additional programmed noise and can operate at a smaller code distance when the intended dissipator is appreciable. This is the resource advantage of partial QEC: target-aligned noise is not treated as failure. The same panel also shows the cost of this advantage. When the assumed physical noise model is mismatched, Strategy A inherits a residual channel error because the compiler has intentionally used physical noise as a resource. Strategy B, by contrast, first suppresses the physical noise and then injects the logical CPTP map through an additional programming step. It is therefore less sensitive to model mismatch, but it cannot use target dissipation to relax the correction requirement. Figure~\ref{fig:channel_compilation} thus gives a concrete instance of the two messages formalized by Theorem~\ref{thm:reachability_convex} and Proposition~\ref{prop:scaling_rk_search}: arbitrary finite-dimensional open quantum system time steps can be compiled whenever the logical channel library is expressive enough, and the required weights $r_k$ are obtained by a polynomial-size convex optimization over implementable logical channels.

\begin{figure}[t]
\centering
\includegraphics[width=\linewidth]{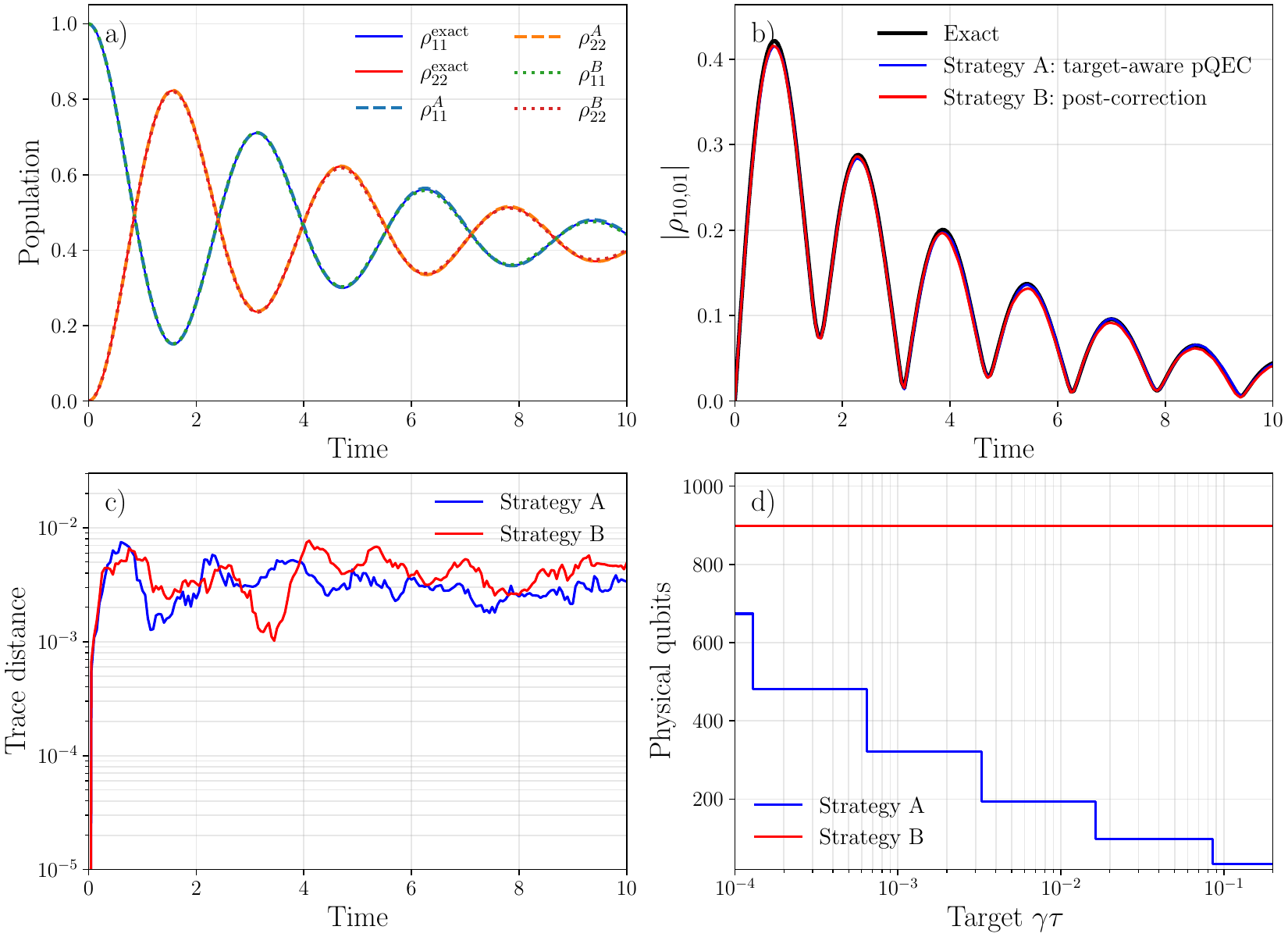}
\caption{\textbf{Open quantum system dynamics and resource trade-off.} 
a) Populations in the two-site exciton model. The exact Lindblad trajectory is compared with a model-aware Strategy-A split step and a post-correction (Strategy B) split step;
b) Coherence $|\rho_{10,01}|$, showing that the programmed dephasing controls the decay envelope;
c) Trace distance from the exact Lindblad trajectory.  The Strategy-A curve includes a deliberate 6\% residual dephasing-rate mismatch to show calibration sensitivity; the Strategy-B curve is dominated by splitting error;
d) Illustrative (surface-code) physical qubits overhead for two logical qubits. Strategy A can use a target-aware threshold proportional to the desired dissipative strength, leading to smaller physical qubits overhead. While Strategy B uses the same overhead as the conventional QEC, the additional logical CPTP programming map mitigates the ancilla qubits for encoding the environments.
}
\label{fig:dynamics_resource}
\end{figure}

Figure~\ref{fig:dynamics_resource} extends the same channel-compilation idea to end-to-end open quantum system dynamics simulations.
Figure~\ref{fig:dynamics_resource}a-b illustrate the excitonic density matrix evolution under exact Lindblad evolution and under the two programmed logical steps. The corresponding error (compared to the Lindbladian dynamics) is plotted in Figure~\ref{fig:dynamics_resource}c. This error captures the combined effects of finite time-step splitting, imperfect channel fitting, residual logical noise and, for Strategy A, mismatch between the calibrated and actual physical noise rates. In the example shown, Strategy A tracks the target dynamics closely when the assumed noise model is accurate, but develops a residual error when a deliberate dephasing-rate mismatch is introduced; Strategy B is insensitive to this calibration error and is instead limited mainly by the accuracy of the logical channel programming and the time-step approximation. Overall, both strategies accurately reproduce the desired dissipative dynamics, capturing the dominant population transfer and decoherence dynamics while maintaining deviations from the Lindblad dynamics below $\sim 1\%$.

The physical qubits needed for the simulation are shown in Figure~\ref{fig:dynamics_resource}d. This resource comparison should be contrasted with conventional dilation-based approaches to open quantum system simulation. At a high level, those methods embed the desired nonunitary channel into a unitary evolution on an enlarged Hilbert space consisting of the system plus auxiliary bath or environment degrees of freedom. From Stinespring, Kraus or singular-value decompositions, the minimal environment dimension is set by the rank needed to represent the channel; for a generic channel acting on a system Hilbert space of dimension $L$, this rank can be as large as $L^2$. Thus, if the system itself requires $n=\log_2 L$ logical qubits, an exact dilation may require up to $2n$ additional logical qubits for the environment, giving a total logical register as large as $3n$ before accounting for controlled-unitary synthesis, reset, measurement, routing and fault-tolerant overheads. In practice, the exact wavefunction of the full system--bath composite is rarely known, and compressed-bath representations~\cite{Cygorek:2022aa} must approximate both the relevant bath modes and their memory effects; even after compression, the auxiliary bath register can remain comparable to or larger than the system register. Partial QEC offers a different route: instead of representing the bath explicitly, it uses calibrated noise, randomized recovery, and logical channel programming to generate the desired dissipative dynamics directly at the logical level.

{\bf Summary and perspective}.
Partial QEC provides a programmable interface between FTQC hardware and open quantum system dynamics. A QEC cycle, together with recovery randomization and logical frame updates, defines a controllable, programmable logical CPTP map. We prove that when the available logical-channel library spans the target dissipator, the desired nonunitary time step can be synthesized by a convex mixture of implementable logical channels. In Strategy A, or model-aware QEC, calibrated hardware noise that is compatible with the target dissipator is deliberately retained at the logical level, so that the code distance is set by the residual wrong channel rather than by the total logical noise. In Strategy B, conventional QEC is first used to suppress circuit errors without relying on a calibrated physical noise model, after which a sampled logical CPTP map implements the desired nonunitary dynamics without explicit bath registers or Stinespring dilations~\cite{Stinespring1955}. The resulting compilation problem is again convex in logical Choi space, making the search for programming probabilities efficient in the size of the logical-channel library rather than in the exponentially large physical Hilbert space of the code block.

This advance changes the logic of early FTQC simulation and suggests several possible routes toward quantum utility in quantum dynamics. Rather than demanding that all logical noise be removed before useful dynamics can be simulated, the partial-QEC framework proposed here seeks to reshape the error space of redundant physical qubits to mimic the desired physical model and to correct only the noise that lies outside that model. This perspective can either lower code-distance requirements or reduce the number of ancilla qubits needed to encode dissipative primitives. Our numerical examples suggest that dephasing, relaxation, transport and non-equilibrium dynamics in open quantum systems may therefore be natural candidates for early quantum utility in the early FTQC era: useful quantum simulations may be achievable not by eliminating noise completely, but by converting calibrated noise and partial correction into programmable logical channels.

Moreover, though this work focuses on open quantum system dynamics, the framework should be viewed more broadly as a method for logical dissipative engineering. Any quantum task that can be formulated as repeated application of a target CPTP map can be a natural applications for partial-QEC, such as dissipative state preparation and ground-state cooling~\cite{Zhan:2026aa, Verstraete:2009aa}.

{\bf Acknowledgements} \\
We acknowledge support from the Laboratory Directed Research and Development (LDRD) program of Los Alamos National Laboratory (LANL).  LANL is operated by Triad National Security, LLC, for the National Nuclear Security Administration of the U.S. Department of Energy (contract no. 89233218CNA000001).

\bibliography{qec, qc4oqs}

\end{document}